\documentclass[aps,pra,superscriptaddress,twocolumn]{revtex4-2}
\usepackage{amsfonts}
\usepackage{amsmath}
\usepackage{mathrsfs}
\usepackage{amssymb}
\usepackage{graphicx}
\usepackage{lipsum} 
\usepackage{bm}
\usepackage{color}
\usepackage{subfigure}

\usepackage[colorlinks=true,linkcolor=blue,anchorcolor=blue,citecolor=blue,urlcolor=blue]{hyperref}

\usepackage{amssymb}
\usepackage{amsmath}
\usepackage{cases}
\usepackage{CJK}
\usepackage{subfigure,graphicx}
\usepackage{dcolumn}
\usepackage{bm}

\begin{document}

\title{Phase diagrams of spin-2 Floquet spinor Bose-Einstein condensates}
\author{Yan-Ling Pan}
\affiliation{College of Physics and Electronic Information Engineering, Qinghai Normal
University, Xining, Qinghai 810016, China}
\author{Qi Li}
\affiliation{College of Physics and Electronic Information Engineering, Qinghai Normal
University, Xining, Qinghai 810016, China}
\author{Gong-Ping Zheng}
\email{Corresponding author: zhenggongping@aliyun.com}
\affiliation{College of Physics and Electronic Information Engineering, Qinghai Normal
University, Xining, Qinghai 810016, China}
\affiliation{Academy of Plateau Science and Sustainability, Xining, Qinghai
810016, China}
\author{Yongping Zhang}                           \email{Corresponding author: yongping11@t.shu.edu.cn}
\affiliation{College of Physics and Electronic Information Engineering, Qinghai Normal
University, Xining, Qinghai 810016, China}
\affiliation{Academy of Plateau Science and Sustainability, Xining, Qinghai
810016, China}
\affiliation{Institute for Quantum Science and Technology, Department of Physics,  Shanghai University, Shanghai 200444, China}

\begin{abstract}
We propose the realization of a spin-2 Floquet spinor Bose-Einstein condensate via Floquet engineering of the quadratic Zeeman energy. In the Floquet system, the coupling strengths of all angular-momentum-conserving spin-flip processes are renormalized by driving-parameter-dependent Bessel functions. Such Floquet-engineered interactions significantly enrich possible states in homogeneous gases. We search for possible states using variational method.  Ground-state phase diagrams, which map the distributions of these possible states, are presented in the space of the driving parameters.

\end{abstract}

\maketitle

\section{Introduction}
Soon after the initial experimental realizations of a scalar Bose-Einstein condensate (BEC), spinor BECs immediately attracted significant experimental and theoretical interest due to their richer degrees of  freedom compared to scalar ones~\cite{Ueda2,yuki12}.  They provide fundamental platforms for studying interacting many-body physics with spin. A spin-1 spinor BEC exhibits both density-density and spin-dependent interactions, with the sign of the latter determining ground-state phase, which can be polar or ferromagnetic~\cite{Ho,Ohmi}. The presence of an external magnetic field can introduce a broken-axisymmetry phase in spin-1 BECs with ferromagnetic interactions (i.e., with a negative spin-dependent interaction sign)~\cite{Murata}. The ground states, spin structures and dynamics have been experimentally studied in various spin-1 BEC systems~\cite{Stenger,Chang2004,Liu2009}. In experiments, the quadratic Zeeman energy can be tuned via a microwave-frequency magnetic field~\cite{Gerbier2006,Leslie2009}. Quantum quenches of this parameter can trigger quantum phase transitions, leading to rich non-equilibrium phenomena~\cite{vinit11,Anquez2016,Yang2019}. Furthermore, merging two distinct spin-1 spinor BECs to form a quantum mixture has been shown to yield a complex and physically meaningful phase diagram~\cite{Jie2010,Shi2010,Xu2010,Li2015}.
Extending the system to the spin-2 spinor BEC reveals even greater complexity, as it possesses two distinct spin-dependent interactions~\cite{Gorlitz2003,Schmaljohann2004,Klempt2009}. This results in phase diagrams that include, in addition to polar and ferromagnetic phases, a cyclic phase~\cite{ho00,Ueda2002,zheng10, yuki11,kob21,kat21}. The spin-2 system is also a promising platform for generating quantum entanglement~\cite{Shi2006,jen15} and studying topological order parameters~\cite{tur07,baio24}. The study of mixtures combining spin-2 and spin-1 spinors is also highly interesting~\cite{eto18,Le2024}.

The fundamental physics of spinor BECs is largely governed by the interplay of density and spin-dependent interactions. However, the experimental tunability of these interaction strengths remains a challenge~\cite{wi06}. In a notable recent study, Fujimoto and Uchino proposed a method to achieve such tunability via Floquet engineering of the quadratic Zeeman energy in a spin-1 spinor BEC~\cite{fuji19}. In a spin-1 system, the spin-dependent interactions include a spin-flip process, 
$|0 \rangle + |0 \rangle \leftrightarrow |-1 \rangle  +  |1 \rangle $, as only this process conserves angular momentum.  This process has attracted significant research interest~\cite{Chang2004,Schmaljohann2004,Wenxian2005,Jacob2012,Pechkis2013,Guan2025}.  It is noteworthy that in their proposed Floquet-engineered system, the high-frequency modulation of the quadratic Zeeman energy specifically modifies the coupling strength of this particular spin-flip process. This modification leads to substantial changes in the possible ground states and phase diagram for ferromagnetic spin-dependent interactions~\cite{fuji19}. More recently, this Floquet engineering approach has been applied to study stripe supersolid phases in spin-1 spinor BECs with experimentally realizable spin-orbit coupling~\cite{zhang23}.

In this work, we apply the Floquet engineering of the quadratic Zeeman energy to a spin-2 spinor BEC. Owing to the greater number of spin degrees of freedom, two distinct spin-dependent interactions in spin-2 condensates possess multiple spin-flip processes that conserve angular momentum~\cite{yuki12}.  Within the high-frequency approximation for the periodically driven quadratic Zeeman energy, we find that the coupling strengths of all these spin-flip processes are modulated by Bessel functions of the first kind. Notably, different processes are governed by different Bessel functions, depending on the specific driving parameters. Consequently, Floquet engineering not only tunes the strengths of these spin-flip interactions but also alters their relative magnitudes. Given the fundamental role of these interactions, this Floquet engineering approach provides a powerful means for the quantum manipulation of spin-2 condensates. To illustrate the difference with a conventional spin-2 condensate, we demonstrate that the resulting Floquet system can be interpreted as possessing the same basic interactions as a conventional one, but with two additional, effective spin-flip interactions derived from the original two spin-dependent interactions.

The introduction of these two Floquet-engineered interactions significantly enriches possible states in homogeneous spin-2 Floquet spinor BECs. We search for possible states using the variational method. They are classified as polar, ferromagnetic, cyclic and broken-axisymmetry phases.
The Bessel-function modulations can alter the energies of these states or directly participate in the construction of their spinor wavefunctions. Notably, we identify novel states that are absent in conventional spinor BECs. Finally, by choosing the lowest-energy state, we map the distribution of these possible states onto a ground-state phase diagram in the space of the driving parameters.

The remainder of the paper is organized as follows. 
In Sec.~\ref{model}, we present the model for a 
spin-2 spinor BEC with a modulated quadratic Zeeman energy. Using the high-frequency approximation, we derive an effective Hamiltonian to describe the resulting Floquet system. In Sec.~\ref{groundstates}, 
we identify possible states of the spin-2 Floquet spinor BEC, including the polar, ferromagnetic, cyclic and broken-axisymmetry phases. In Sec.~\ref{diagram}, corresponding ground-state phase diagrams are presented in the parameter space of the driving field. A summary of our findings is provided in Sec.~\ref{conclusion}. Finally, the Appendix presents an alternative derivation of the same effective Hamiltonian using the high-frequency expansion method.

\section{Spin-2 Floquet spinor Bose condensates}
\label{model}

The spin-2 spinor BEC features a vector field operator $\hat{\psi}=\left( \hat{\psi}_{2},\hat{\psi}_{1},\hat{\psi}_{0},\hat{%
\psi}_{-1},\hat{\psi}_{-2}\right) ^{T}$, with $\hat{\psi}_{m_{F}}$ representing the wave function of  Zeeman states $ | F=2, m_{F} \rangle \equiv | m_{F} \rangle $ ($m_{F}=\{2,1,0,-1,-2 \}  $) within a spin $F=2$ manifold.  The BEC system is described by the Hamiltonian~\cite{yuki12}, 
\begin{align}
\hat{H}= & \int d\mathbf{r} \hat{\psi}^{\dagger }\left( -\frac{\hbar 
^{2}\triangledown ^{2}}{2M}\right) \hat{\psi} + \int d\mathbf{r} \hat{\psi}^{\dagger }\left( qF_{z}^{2}\right) \hat{\psi} \notag \\
&+  \frac{c_{0}}{2}\int d\mathbf{r}\left( \hat{\psi}^{\dagger }%
\hat{\psi}\right) ^{2}  
+\frac{c_{1}}{2}\int d\mathbf{r}\left( \hat{\psi}^{\dagger }%
\mathbf{{F}}\hat{\psi}\right) \cdot \left( \hat{\psi}^{\dagger }\mathbf{%
{F}}\hat{\psi}\right)  \notag \\
&+
\frac{c_{2}}{2}\int d\mathbf{r}\left( \hat{\psi}^{T}{A%
}\hat{\psi}\right) ^{\dagger }\left( \hat{\psi}^{T}{A}\hat{\psi}%
\right).
\label{Orignal}
\end{align}
Here, the first term represents the kinetic energy,
where $M$  is the atomic mass.  
\begin{equation}
 \mathbf{{F}}
=(F_{x},F_{y},F_{z}),
\end{equation}
  and it denotes the spin-$2$ Pauli matrices~\cite{yuki12}. The second term is the effective quadratic Zeeman energy, with $q$ being its magnitude. The remaining three terms describe the interactions:  $c_{0}$ is the spin-independent interaction coefficient, representing the density-density interaction; $c_{1}$ and $c_{2}$ are spin-dependent interaction coefficients describing the spin-spin interaction and interaction through the spin-singlet scattering channel respectively.  ${A}$ is
associated with the amplitude of the spin-singlet pair \cite{ho00}, 
\begin{equation}
    A= \begin{pmatrix}
        0 & 0&  0& 0& 1 \\
        0 & 0& 0  & -1 & 0 \\
        0 & 0 & 1 & 0 &0  \\
        0 & -1 & 0 & 0 &0 \\
        1 & 0 & 0 & 0 &0
    \end{pmatrix},
\end{equation}
which is an antisymmetric matrix. The field operator of the spin-singlet state becomes $ \hat{\psi}^{T}{A%
}\hat{\psi}$ with $T$ being transpose operator.

In experiments, the quadratic Zeeman shift is tunable~\cite{Gerbier2006,Leslie2009,vinit11,Anquez2016,Yang2019}. It can be experimentally adjusted by applying a linearly polarized microwave-frequency magnetic field which off-resonantly induces the transition between the $ | F=2, m_{F}=0 \rangle $ and $ | F=1, m_{F}=0 \rangle $ hyperfine states. The transition, which has the coupling strength $\Omega$, is detuned by $\delta $. The quadratic Zeeman shift $q$ is proportional to $\hbar \Omega^2/\delta$~\cite{Leslie2009}. By periodically modulating the strength $\Omega$ (via the modulation of the magnitude of the magnetic field), we implement the quadratic Zeeman shift as~\cite{fuji19}
\begin{equation}
q=q_0 +Q \cos (\omega t),
\end{equation}
where $\omega $ and $Q$ are the driving frequency and amplitude, respectively, and $q_0$ is the averaged oscillation value. We assume the modulation is high-frequency, i.e., the driving frequency is much larger than other typical energy scales of the system.

Under the high-frequency assumption, the spinor system can be described in a time-independent way. This can be carried out by performing a unitary transformation to cancel the time dependence $Q\cos(\omega t)$ and by averaging the resultant Hamiltonian over oscillating periods~\cite{eck15}. The time-dependent unitary transformation is $\hat{\psi}= U\hat{\phi}$ with the new field operator 
$\hat{\phi}=\left( \hat{\phi}_{2},\hat{\phi}_{1},\hat{\phi}_{0},\hat{%
\phi}_{-1},\hat{\phi}_{-2}\right) ^{T}$, and $U=\exp\left( -i\mathcal{Q}\sin (\omega t)F_{z}^{2}\right) $
with
\begin{equation}
\mathcal{Q}= \frac{Q}{\hbar \omega }.
\end{equation}
Substituting the transformation into $\hat{H}-i\hbar \int d \mathbf{r} \hat{\psi}^\dagger \partial \hat{\psi}/\partial t$, we obtain 
\begin{align}
\hat{H}' =& \int d\mathbf{r} \hat{\phi}^{\dagger }\left( -\frac{\hbar 
^{2}\triangledown ^{2}}{2M}\right) \hat{\phi} + \int d\mathbf{r} \hat{\phi}^{\dagger }\left( q_0F_{z}^{2}\right) \hat{\phi} \notag \\
& +  \frac{c_{0}}{2}\int d\mathbf{r}\left( \hat{\phi}^{\dagger }%
\hat{\phi}\right) ^{2} +H_\text{sd}'(t,\hat{\phi}).
\end{align}
The resultant spin-dependent interactions $H_\text{sd}'(t,\hat{\phi})$ become very complicate and time-dependent in the way of the presence of factors $\exp[\pm i j \mathcal{Q} \sin(\omega t) ]$ (with $j$ an integer) in front of the coupling strengths in all spin-flip processes. Nevertheless,  $H_\text{sd}'(t,\hat{\phi})$ can be greatly simplified after performing the time average. Physically, the unitary transformation may be considered as displacing the system into the fast oscillating frame of the quadratic Zeeman shift. In this frame, the system is stationary. This can be achieved by performing the time average to $\hat{H}' $, i.e., $ \hat{H}_\text{eff} =  \omega/(2\pi)\int_0^{2\pi/\omega }dt  \hat{H}'  $~\cite{eck15}, Finally, we obtain the effective Hamiltonian,
\begin{widetext}
\begin{align}
\label{effectiveH}
\hat{H}_\text{eff} = & \int d\mathbf{r} \hat{\phi}^{\dagger }\left( -\frac{\hbar 
	^{2}\triangledown ^{2}}{2M}\right) \hat{\phi} + \int d\mathbf{r} \hat{\phi}^{\dagger }\left( q_0F_{z}^{2}\right) \hat{\phi}  
+ \frac{c_{0}}{2}\int d\mathbf{r}\left( \hat{\phi}^{\dagger }%
\hat{\phi}\right) ^{2}  +\frac{c_{1}}{2}\int d\mathbf{r}\left( \hat{\phi}^{\dagger }%
\mathbf{{F}}\hat{\phi}\right) \cdot \left( \hat{\phi}^{\dagger }\mathbf{%
	{F}}\hat{\phi}\right)  \notag \\
&+ 
\frac{c_{2}}{2}\int d\mathbf{r}\left( \hat{\phi}^{T}{A%
}\hat{\phi}\right) ^{\dagger }\left( \hat{\phi}^{T}{A}\hat{\phi}%
\right) + \hat{H}_\text{sf}, \\
\hat{H}_\text{sf} = & \frac{c_1}{2} \int d\mathbf{r}  \left\{ \left[ J_0(2\mathcal{Q})-1 \right] \left( 6\hat{\phi}
_{0}^{\dagger }\hat{\phi}_{0}^{\dagger }\hat{\phi}_{1}\hat{\phi}_{-1} +  2\sqrt{6
}\hat{\phi}_{1}^{\dagger }\hat{\phi}_{1}^{\dagger }\hat{\phi}_{2}\hat{\phi}%
_{0}  +2\sqrt{6}\hat{\phi}_{-1}^{\dagger }\hat{\phi}%
_{-1}^{\dagger }\hat{\phi}_{0}\hat{\phi}_{-2} \right)  +4\left[ J_0(6\mathcal{Q})-1\right] \hat{\phi}_{-1}^{\dagger }\hat{\phi}_{1}^{\dagger }\hat{\phi}_{2}\hat{%
\phi}_{-2} \right. \notag \\
&\left. +2\sqrt{6} \left[  J_0(4\mathcal{Q})-1\right] \left(\hat{\phi}_{-1}^{\dagger }\hat{\phi}_{0}^{\dagger }\hat{%
\phi}_{1}\hat{\phi}_{-2} +\hat{\phi}_{0}^{\dagger }\hat{\phi}_{1}^{\dagger }\hat{%
\phi}_{2}\hat{\phi}_{-1} \right)+\text{H.c.}\right\} +\frac{c_2}{2}\int d\mathbf{r}\left\{ 2 \left[ J_0(8\mathcal{Q})-1 \right]  \hat{\phi}_{0}^{\dagger }\hat{%
\phi}_{0}^{\dagger }\hat{\phi}_{2}\hat{\phi}_{-2}     \right. \notag \\
& \left.  -4\left[ J_0(6\mathcal{Q})-1 \right] \hat{\phi}_{-1}^{\dagger }\hat{\phi}%
_{1}^{\dagger }\hat{\phi}_{2}\hat{\phi}_{-2}- 2 \left[ J_0(2\mathcal{Q})-1 \right] \hat{\phi}_{0}^{\dagger }\hat{\phi}_{0}^{\dagger }\hat{\phi}_{1}%
\hat{\phi}_{-1}+ \text{H.c.}\right\}.  \notag
\end{align}
\end{widetext}
In comparison with the original Hamiltonian $\hat{H}$, the Floquet system $\hat{H}_\text{eff}$ possesses additional Floquet engineered spin-flip couplings $\hat{H}_\text{sf}$.
During deriving the effective Hamiltonian, we already use $\exp[\pm i j\mathcal{Q}\sin(\omega t)]= \sum_n J_n(j\mathcal{Q} )\exp(\pm i n \omega t)$ with $J_n$ the n-th order Bessel function of the first kind. Only $J_0$ (i.e., n=0) are retained in $\hat{H}_\text{sf}$ due to the time average.

Following the convention by Fujimoto and Uchino~\cite{fuji19}, we name the Floquet engineered $\hat{H}_\text{eff}$ as spin-2 Floquet spinor BECs. In spin-1 systems, the difference between usual and Floquet spinor BECs lies in that the Floquet one has an extra spin-exchange process of $|0 \rangle + |0 \rangle \leftrightarrow |-1 \rangle  +  |1 \rangle $, which is the only spin-flip coupling conserving angular momentum~\cite{fuji19,zhang23}. Similarly, the spin-2 Floquet spinor BEC possesses extra spin-flip interacting couplings represented by $\hat{H}_\text{sf}$.  For a spin-2 spinor BEC, the spin-spin interaction, $( \hat{\psi}^{\dagger }%
\mathbf{{F}}\hat{\psi}) \cdot (\hat{\psi}^{\dagger }\mathbf{%
{F}}\hat{\psi} )$, is composed of density couplings such as $\hat{\psi}_1^\dagger\hat{\psi}_2^\dagger \hat{\psi}_2 \hat{\psi}_1$ and spin-flip couplings of $|m_1 \rangle + |m_2 \rangle \leftrightarrow |m_1+1 \rangle  +  |m_2-1 \rangle $ (with $(m_1,m_2)=\{(1,1),(0,0),(-1,-1),(1,-1),(0,-1),(1,0) \}$)~\cite{yuki12}.  The Floquet engineering of the quadratic Zeeman energy manipulates all these spin-flip couplings by modulating their relative strengths via Bessel functions. Take the spin-flip process $ \hat{\psi}_{-1}^\dagger \hat{\psi}_0^\dagger \hat{\psi}_1 \hat{\psi}_{-2}  $ as an example: 
\begin{align}
 \label{spinflipexample}
 &\hat{\psi}_{-1}^\dagger \hat{\psi}_0^\dagger \hat{\psi}_1 \hat{\psi}_{-2}  \notag \\
 &\rightarrow   \int dt e^{-i4\mathcal{Q} \sin(\omega t)}  \hat{\phi}_{-1}^\dagger \hat{\phi}_0^\dagger \hat{\phi}_1 \hat{\phi}_{-2} =J_0 ( 4\mathcal{Q}  ) \hat{\phi}_{-1}^\dagger \hat{\phi}_0^\dagger \hat{\phi}_1\hat{\phi}_{-2}
 \notag  \\
 &=\hat{\phi}_{-1}^\dagger \hat{\phi}_0^\dagger \hat{\phi}_1\hat{\phi}_{-2}  +\left[J_0 ( 4\mathcal{Q}  ) -1\right] \hat{\phi}_{-1}^\dagger \hat{\phi}_0^\dagger\hat{\phi}_1\hat{\phi}_{-2}.
 \end{align}
Therefore, it seems that the Floquet engineering generates a prefactor $J_0 ( 4\mathcal{Q}  ) $ in front of the process  $\hat{\phi}_{-1}^\dagger \hat{\phi}_0^\dagger \hat{\phi}_1\hat{\phi}_{-2}$.  The engineering does not modify density couplings since there is no time-dependent phase involved after the transformation. Take the density coupling $\hat{\psi}_{1}^\dagger \hat{\psi}_2^\dagger \hat{\psi}_2 \hat{\psi}_{1}$ as an example: 
\begin{align}
 \hat{\psi}_{1}^\dagger \hat{\psi}_{2}^\dagger \hat{\psi}_2 \hat{\psi}_{1} \rightarrow  & \int dt \hat{\phi}_{1}^\dagger \hat{\phi}_2^\dagger \hat{\phi}_2 \hat{\phi}_{1} =\hat{\phi}_{1}^\dagger \hat{\phi}_2^\dagger \hat{\phi}_2\hat{\phi}_{1}.
 \label{example}
 \end{align}
All these density couplings together with spin-flip couplings as the first term in the last equation in Eq.~(\ref{spinflipexample}) are re-expressed as  $( \hat{\phi}^{\dagger }%
\mathbf{{F}}\hat{\phi}) \cdot (\hat{\phi}^{\dagger }\mathbf{%
{F}}\hat{\phi} )$ which appears in Eq.~(\ref{effectiveH}). 
The Bessel-function modulated spin-flip terms, such as the one with coefficient $J_0 ( 4\mathcal{Q} )-1$ in the last equation of  Eq.~(\ref{spinflipexample}), are grouped under the term governed by the coefficient $c_1/2$ in  $\hat{H}_\text{sf}$.  In the same way, the Floquet engineering also modulates the spin-flip processes in the spin-singlet channel which are $|0 \rangle + |0 \rangle \leftrightarrow |1 \rangle  +  |-1 \rangle $,  $|0 \rangle + |0 \rangle \leftrightarrow |2 \rangle  +  |-2 \rangle $, and  $|1 \rangle  +|-1\rangle\leftrightarrow |2 \rangle  +  |-2 \rangle $. These are gathered and described by the term with coefficient $c_2/2$ in  $\hat{H}_\text{sf}$.  The minus signs in front of the coupling strengths in some terms originate from a minus sign in the field operator of the spin-singlet pair 
$ \hat{\psi}^{T}{A}\hat{\psi}=  2 \hat{\psi}_2\hat{\psi}_{-2} -2\hat{\psi}_1\hat{\psi}_{-1}  +\hat{\psi}_0\hat{\psi}_{0}$.

Floquet engineering of the quadratic Zeeman energy provides an experimentally accessible method for tuning the interactions in a spin-2 spinor BEC. This is possible because the quadratic Zeeman term does not commute with the interaction terms in the Hamiltonian. We derive the corresponding Floquet-engineered effective Hamiltonian, $\hat{H}_\text{eff}$ in Eq.~(\ref{effectiveH}) using the high-frequency expansion method~\cite{fuji19,gold14}; detailed calculations are provided in the appendix. The derivation clearly shows that this non-commutation leads to the non-trivial modulation of interactions. It is well-known that the interactions in spin-2 spinor BECs possess SO(3) symmetry in spin space. Consequently, any unitary transformation generated by spin operators, such as  $\exp(i{F}_z)$ or $\exp(i{F}_x)$, leaves the interaction Hamiltonian invariant. This explains why Floquet engineering of typical experimental parameters like the linear Zeeman energy ($\sim{F}_z$) or a Rabi coupling ($\sim{F}_x$) cannot tune the interactions, in contrast to the quadratic Zeeman energy.

\section{ Possible states   }
\label{groundstates}

The ground state of a uniform spin-2 spinor BEC can be the polar, ferromagnetic and cyclic phases~\cite{yuki12,ho00,Ueda2002,zheng10}. Due to the Bessel-function modulations of all spin-flip interacting couplings introduced by Floquet engineering, it is crucial to investigate how  possible phases and ground-state phase diagrams are modified in a spin-2 Floquet spinor BEC.

Possible states are analyzed using the mean-field theory. The mean-field approximation replaces field operators 
by their averaged values, $ \hat{\phi} \rightarrow \langle \hat{\phi} \rangle = \phi$, which become wave functions. Since the system is homogeneous, they are  assumed to be 
$\phi= \sqrt{n}\zeta$, where $n=N/V$ is the density with $N$ the total atom number and $V$ the size of the system and $ |\zeta\rangle =( \zeta_2,\zeta_1,\zeta_0,\zeta_{-1},\zeta_{-2} )^T$ is the normalized spinor with  $\langle \zeta | \zeta \rangle =1$.  Under the mean-field approximation, the energy functional associated with the effective Hamiltonian in Eq.~(\ref{effectiveH}) becomes,
\begin{equation}
\label{Energyfunctional}
\frac{E }{N} =q_0 \langle
F_{z}^{2}\rangle+\frac{\Tilde{c}_{0} }{2} +\frac{\Tilde{c}_{1}}{2}\langle \mathbf{F}\rangle ^{2}  +\frac{\Tilde{c}_{2}}{2}
\left\vert \Theta \right\vert ^{2} +E_{F},
\end{equation}
with the Floquet engineered nonlinearity  
\begin{align}
\label{engineernonlinearity}
& E_F = 
\frac{\Tilde{c}_1}{2}\biggl\{\left[J_0(2\mathcal{Q})-1\right]\bigl(6\zeta_{0}^{\ast}\zeta_{0}^{\ast}\zeta_{1}\zeta_{-1} +2\sqrt{6}\zeta_{1}^{\ast}\zeta_{1}^{\ast}\zeta_{2}\zeta_{0} \notag\\
&+2\sqrt{6}\zeta_{-1}^{\ast}\zeta_{-1}^{\ast}\zeta_{0}\zeta_{-2}\bigr) +4\left[J_0(6\mathcal{Q})-1\right] \zeta _{-1}^{\ast }\zeta _{1}^{\ast }\zeta _{2}\zeta _{-2} \notag\\
&+2\sqrt{6}\left[J_0(4\mathcal{Q})-1\right]
\bigl(\zeta _{-1}^{\ast }\zeta _{0}^{\ast }\zeta _{1}\zeta _{-2}+\zeta _{0}^{\ast }\zeta _{1}^{\ast }\zeta _{2}\zeta _{-1}\bigr)+\text{c.c.}\biggr\} \notag \\
& +\frac{\Tilde{c}_2}{2}\biggl\{ 2\left[J_0(8\mathcal{Q})-1\right]\zeta _{0}^{\ast }\zeta _{0}^{\ast }\zeta _{2}\zeta _{-2} -4\left[J_0(6\mathcal{Q})-1\right] \notag \\
&\times \zeta _{-1}^{\ast }\zeta _{1}^{\ast }\zeta _{2}\zeta
_{-2}
-2\left[J_0(2\mathcal{Q})-1\right]\zeta _{0}^{\ast }\zeta _{0}^{\ast }\zeta _{1}\zeta _{-1}+\text{
c.c.}\biggr\}.
\end{align}
The interacting coefficients are rescaled as $ \Tilde{c}_0=c_0N$, $ \Tilde{c}_1=c_1N$ and $ \Tilde{c}_2=c_2N$.
In Eq.~(\ref{Energyfunctional}), there are three order parameters, $\langle
F_{z}^{2}\rangle=\langle \zeta | F_{z}^{2} | \zeta \rangle$,  $\langle \mathbf{F}\rangle= \langle \zeta | \mathbf{F} | \zeta \rangle$, and $\Theta= (| \zeta \rangle)^T {A} |  \zeta \rangle =2\zeta _{2}\zeta
_{-2}-2\zeta _{1}\zeta _{-1}+\zeta _{0}^{2}$. 
Due to $e^{i\theta F_z} \hat{H}_\text{eff} e^{-i\theta F_z} = \hat{H}_\text{eff}$ with $\theta$ being an arbitrary angle,  the longitudinal magnetization $\langle F_{z}\rangle$ is conserved, and we set its value as  $\langle F_{z}\rangle=m$. The transverse magnetization is defined as $\langle 
{F}_{+}\rangle$ with ${F}_{+} = F_x+ iF_y $.

The spin configuration of the spinor  $|\zeta \rangle$ is presumed by a variational ansatz. Variational parameters 
are determined by minimizing the energy
functional under several constraints~\cite{Wenxian2003},
such as, the normalization condition and the conservation of $\langle
F_{z}\rangle$, which are respectively listed as follows, 
\begin{align}
\label{constraints}
 & \langle \zeta | \zeta \rangle= \left\vert \zeta _{2}\right\vert ^{2}+\left\vert \zeta _{1}\right\vert
^{2}+\left\vert \zeta _{0}\right\vert ^{2}+\left\vert \zeta _{-1}\right\vert
^{2}+\left\vert \zeta _{-2}\right\vert ^{2}=1, \notag \\
&\langle {F}_{z}\rangle =2\left\vert \zeta _{2}\right\vert
^{2}-2\left\vert \zeta _{-2}\right\vert ^{2}+\left\vert \zeta
_{1}\right\vert ^{2}-\left\vert \zeta _{-1}\right\vert ^{2} = m.
\end{align}
Possible states obtained by this variational method are not necessarily ground states of the system. To show ground-state phase diagram, we compare energies of possible states and choose the one having lowest energy. Note that the only requirement for the variational method is that each component shares an identical, homogeneous spatial profile. Consequently, the variational approach remains valid without considering solutions that possess complex spatial structures, such as fragmented states or spin domains. However, external trapping potentials-unavoidable in real atomic gas experiments-introduce spatial inhomogeneity. To ensure that the variational results and the ground-state phase diagram remain applicable, these trapping potentials must be kept sufficiently weak.

According to results of the minimization with the constraints, possible states are identified as polar, ferromagnetic, cyclic and broken-axisymmetry
phases in the following.

\subsection{Polar phase}

The polar phase features a nonzero order parameter of $\Theta$ and a zero transverse magnetization, i.e., $\Theta \neq 0$, and $  \langle {F}_{+}\rangle =2\zeta _{2}^{\ast }\zeta _{1}+%
\sqrt{6}\zeta _{1}^{\ast }\zeta _{0}+\sqrt{6}\zeta _{0}^{\ast }\zeta
_{-1}+2\zeta _{-1}^{\ast }\zeta _{-2}=0$~\cite{ho00}. 
By checking $\Theta =2\zeta _{2}\zeta
_{-2}-2\zeta _{1}\zeta _{-1}+\zeta _{0}^{2}$, there are six possible cases to make $\Theta \neq 0$: (1) $
\zeta _{0}\neq 0$; (2) $\zeta _{1}\neq 0$ and $\zeta _{-1}\neq 0$; (3) $\zeta
_{2}\neq 0$ and $\zeta _{-2}\neq 0$;  (4) $\zeta _{2}\neq 0$, $\zeta _{-2}\neq 0 $, and $\zeta
_{0}\neq 0$; (5) $\zeta _{1}\neq 0$, $\zeta _{-1}\neq 0$, and $\zeta _{0}\neq 0$; (6) all $\zeta _{m_F}\neq 0$. According to these different cases, we build up trial wavefunction ansatz that satisfy the constraints in Eq.~(\ref{constraints}) and $\langle {F}_{+}\rangle =0$~\cite{yuki12, zheng10}.  Using the variational method, the ansatz can be determined. 

(1) When $\zeta _{0}\neq 0$ and $\zeta _{2}=\zeta _{-2}=\zeta _{1}=\zeta _{-1}=0$, we have the polar solution as
\begin{equation}
\mathbf{P_{0}}: \    \left(
\begin{array}{ccccc}
0, & 0, & e^{i\theta _{0}}, & 0, & 0%
\end{array}%
\right) ^{T}.
\end{equation}%
Here $\theta _{0}$ is the U(1) gauge-invariant phase. Its energy is $E_{P_0}/N=\Tilde{c}_{0}/2+
\Tilde{c}_{2}/2$.

(2) When $\zeta _{1}\neq 0$ and $\zeta _{-1}\neq 0$, and $\zeta _{2}=\zeta _{-2}=\zeta _{0}=0$, the polar solution becomes 
\begin{equation}
\mathbf{P_{1}}: \   \left(
\begin{array}{ccccc}
0, & \sqrt{\frac{1+m}{2}}e^{i\theta _{1}}, & 0, & \sqrt{\frac{1-m}{2}}%
e^{i\theta _{-1}}, & 0%
\end{array}%
\right) ^{T},
\end{equation}
with $ \theta _{1}$ and $\theta _{-1}$ being arbitrary angles. Its energy is 
$E_{P_1}/N=q_0+\Tilde{c}_{0}/2+m^2\Tilde{c}_{1}/2+(1-m^2)\Tilde{c}_{2}/2$.

(3) When $\zeta _{2}\neq 0$ and $\zeta _{-2}\neq 0$, and  $\zeta _{1}=\zeta _{-1}=\zeta _{0}=0$, the state is 
\begin{equation}
\mathbf{P_{2}}: \  \left(
\begin{array}{ccccc}
\frac{\sqrt{2+m}}{2}e^{i\theta _{2}}, & 0, & 0, & 0, & \frac{\sqrt{2-m}}{2}%
e^{i\theta _{-2}}%
\end{array}%
\right) ^{T}.
\end{equation}
Again $\theta _{2}$ and $\theta _{-2}$ are arbitrary angles. Its energy becomes $E_{P_2}/N=4q_0+\Tilde{c}_{0}/2+m^2\Tilde{c}_{1}/2+(1-m^2)\Tilde{c}_{2}/2$.

(4) When $\zeta _{2}\neq 0$, $\zeta _{-2}\neq 0$ and $\zeta _{0}\neq 0$, and  $\zeta _{1} = \zeta _{-1}=0$,  we find several solutions if considering $m=0$,
\begin{equation}
\mathbf{P_{3}}: \   \left(
\begin{array}{ccccc}
\sqrt{\frac{1-\beta ^{2}}{2}}e^{i\chi }, & 0, & \beta,
& 0, & \sqrt{\frac{1-\beta ^{2}}{2}}e^{-i\chi }%
\end{array}%
\right) ^{T},
\end{equation}
with $\chi$ being an arbitrary angle, and 
\begin{equation}
    \beta =\sqrt{\frac{4q_0+\Tilde{c}_2-\Tilde{c}_2J_0(8\mathcal{Q})}{2\Tilde{c}_2-2\Tilde{c}_2J_0(8\mathcal{Q})}},
\end{equation}
and the corresponding energy
\begin{equation}
\frac{E_{P_3}}{N}=2q_0+\frac{\Tilde{c}_0}{2}+\frac{\Tilde{c}_2}{4}+\frac{\Tilde{c}_2}{4}J_0(8\mathcal{Q})-\frac{4q_0^2}{\Tilde{c}_2-\Tilde{c}_2J_0(8\mathcal{Q})}.
\end{equation}
\begin{equation}
\mathbf{P_{4}}: \  \left(
\begin{array}{ccccc}
-\sqrt{\frac{1-\gamma ^{2}}{2}}e^{i\chi}, & 0, & \gamma
, & 0, & \sqrt{\frac{1-\gamma ^{2}}{2}}e^{-i\chi}%
\end{array}%
\right) ^{T},
\end{equation}%
with 
\begin{equation}
\gamma =\sqrt{\frac{4q_0+\Tilde{c}_2+\Tilde{c}_2J_0(8\mathcal{Q})}{2\Tilde{c}_2+2\Tilde{c}_2J_0(8\mathcal{Q})}},
\end{equation}
and associated energy
\begin{equation}
\frac{E_{P_4}}{N} =2q_0+\frac{\Tilde{c}_0}{2}+\frac{\Tilde{c}_2}{4}-\frac{\Tilde{c}_2}{4}J_0(8\mathcal{Q})-\frac{4q_0^2}{\Tilde{c}_2+\Tilde{c}_2J_0(8\mathcal{Q})}.
\end{equation}
\begin{equation}
\mathbf{P_{5}}:\   \left(
\begin{array}{c}
\frac{\sqrt{1-\beta ^{2}}-\sqrt{1-\gamma ^{2}}}{\sqrt{2}}e^{i\chi} \\
0 \\
\sqrt{\beta ^{2}+\gamma ^{2}-1} \\
0 \\
\frac{\sqrt{1-\beta ^{2}}+\sqrt{1-\gamma ^{2}}}{\sqrt{2}}e^{-i\chi}%
\end{array}%
\right),
\end{equation}%
and its energy is
\begin{align}
\frac{E_{P_5}}{N} =& 4q_0 (2-\beta ^{2}-\gamma ^{2})+\frac{ \Tilde{c}_{0}}{2}
+8\Tilde{c}_{1}(1-\beta ^{2})(1-\gamma ^{2}) \notag \\
&+\Tilde{c}_{2}[(\beta^4+\gamma^4)-(\beta^2+\gamma^2)]  \notag \\
&+\Tilde{c}_{2}J_0(8\mathcal{Q})[(\gamma ^{4}-\beta ^{4})-(\gamma ^{2}-\beta ^{2})].
\end{align}
Note that in the solutions $\mathbf{P_{3}} $,  $\mathbf{P_{4}}$ and $\mathbf{P_{5}}$ the nonlinear coefficient $\Tilde{c}_2$ not only affects on the energy but also participates to wavefunctions. While there is no $\Tilde{c}_1$ in the wavefunctions, as in the Floquet engineered nonlinear $E_F$ in Eq.~(\ref{engineernonlinearity}) the working of $\Tilde{c}_1$ requires $\zeta_1\neq 0$ or $\zeta_{-1}\neq 0$.

(5) In the case of $\zeta _{2}=\zeta _{-2}=  0$, $\zeta _{1}\neq 0$, $\zeta _{0}\neq 0$ and $\zeta _{-1}\neq 0$, we find polar solution with $m=0$,
\begin{equation}
\mathbf{P_{6}}: \  \left(
\begin{array}{ccccc}
0, & -\sqrt{\frac{1-\alpha ^{2}}{2}}e^{i\chi}, & \alpha
, & \sqrt{\frac{1-\alpha ^{2}}{2}}e^{-i\chi}, & 0%
\end{array}%
\right) ^{T}.
\end{equation}%
with 
\begin{equation}
\alpha =\sqrt{ \frac{1}{2} - \frac{ q_0}{2(3\Tilde{c}_1-\Tilde{c}_2)[1-J_0(2\mathcal{Q})]}},
\label{alpha}
\end{equation}
and its energy is
\begin{align}
\frac{E_{P_6}}{N} =& \frac{q_0}{2}+\frac{ \Tilde{c}_{0}}{2}+\frac{ \Tilde{c}_{2}}{2}+\frac{q
_0^2}{4(3\Tilde{c}_{1}-\Tilde{c}_{2})[1-J_0(2\mathcal{Q}) ]}  \notag \\
&+\frac{(3\Tilde{c}_{1}-\Tilde{c}_{2})[1-J_0{(2\mathcal{Q})}]}{4}.
\end{align}
It has already been shown that this solution cannot exist in a conventional spin-2 condensate~\cite{yuki12}. Our result is consistent with this conclusion: When $\mathcal{Q}=0$, $J_0(2\mathcal{Q})=1$, which leads to the non-definition of $\alpha$ from Eq.~(\ref{alpha}) if $q_0\neq 0$. The interesting finding is that the driving can support its existence, which becomes a unique feature of Floquet spinor BECs. It is noticed that the solutions $\mathbf{P_{3}}$, $\mathbf{P_{4}}$, $\mathbf{P_{5}}$ and $\mathbf{P_{6}}$ have $m=0$. If $m\neq 0$, it is impossible to get analytical expressions for them.

Finally, we discuss case (6), where all $\zeta_{m_F}\neq 0$. In order to have the zero transverse magnetization, the solution in the case (6) should be the superposition of   $\mathbf{P_{2}}$ and $\mathbf{P_{6}}$. However, since $\mathbf{P_{2}}$ and $\mathbf{P_{6}}$ have different chemical potential, the superposition can not lead to a stationary ground state.  

\subsection{Ferromagnetic phase}

The ferromagnetic phase belongs to eigenstates of $F_{z}$. There are the following possible states:
\begin{equation}
\mathbf{F_{1}}: \  \left(
\begin{array}{ccccc}
0, & e^{i\theta _{1}}, & 0, & 0, & 0%
\end{array}%
\right) ^{T}, 
\end{equation}%
with the energy $E_{F_1}/N=q_0+\Tilde{c}_{0}/2+\Tilde{c}_{1}/2$, and
\begin{equation}
\mathbf{F_{2}}: \  \left(
\begin{array}{ccccc}
e^{i\theta _{2}}, & 0, & 0, & 0, & 0%
\end{array}%
\right) ^{T},
\end{equation}
with the energy $E_{F_2}/N=4q_0+\Tilde{c}_{0}/2+2\Tilde{c}_{1}$.

\begin{figure*}[t]
\centering
{\includegraphics[width =1\textwidth]{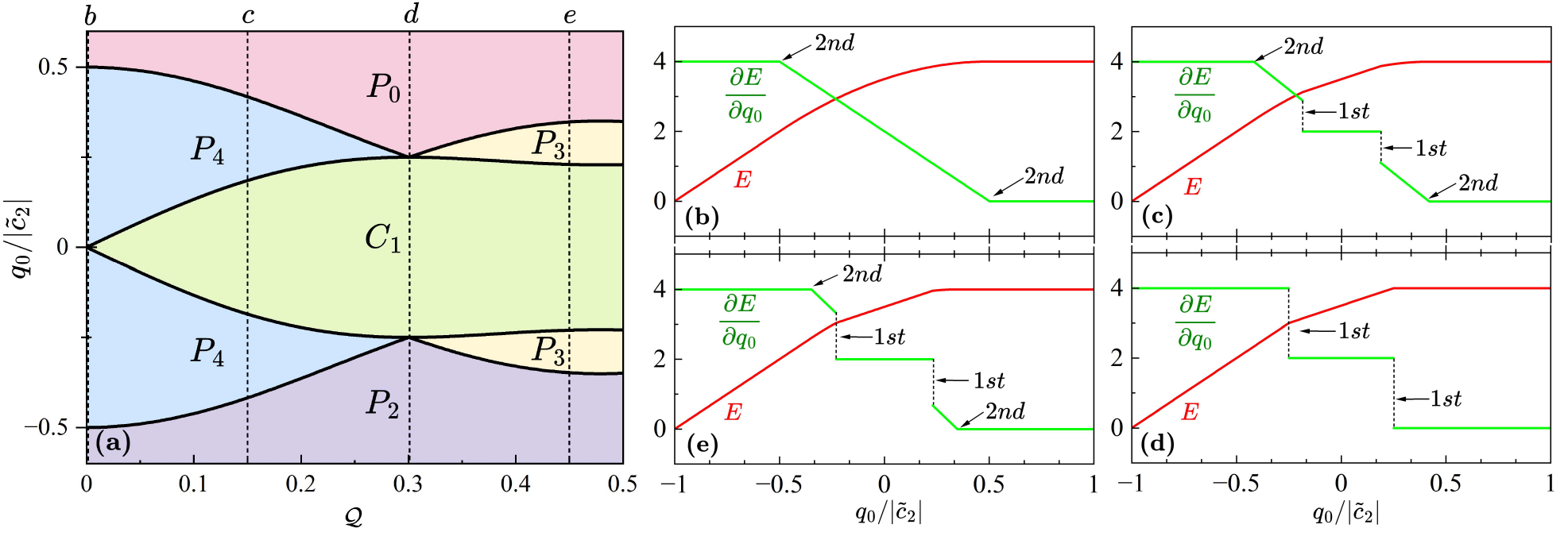}}
\caption {Ground-state phase diagram  for the anti-ferromagnetism $\Tilde{c}_1>0$ with $\Tilde{c}_2>0$ and $\langle F_{z}\rangle=m=0$. (a) Diagram in the parameter space of periodic driving $(q_0,\mathcal{Q}=Q/ (\hbar \omega))$.  Vertical dashed lines labeled by b,c,d,e correspond to $\mathcal{Q}=( 0, 0.15, 0.3006,0.45)$ respectively. The energy $E$ (red lines) and derivative of energy $\partial E/\partial q_0$ (green lines) along these labeled lines are demonstrated in (b) ($\mathcal{Q}=0$), (c) ($\mathcal{Q}=0.15$), (d) ($\mathcal{Q}=0.3006$) and
(e) ($\mathcal{Q}=0.45$), respectively.  The
unit of energy is $%
\left\vert \Tilde{c}_{2}\right\vert $. Other parameters are $ \Tilde{c}_{0}=7\left\vert
\Tilde{c}_{2}\right\vert $ and $\Tilde{c}_1=0.5\left\vert
\Tilde{c}_{2}\right\vert $.}
\label{fig1}
\end{figure*}

\subsection{Cyclic phase}

The cyclic phase features $\Theta =0$ and $\langle {F}_{+}\rangle =0$~\cite{ho00,Ueda2002,zheng10}. We use the same approach for the previous study of polar phases to construct cyclic wavefunctions considering a more constraint $\Theta =2\zeta _{2}\zeta
_{-2}-2\zeta _{1}\zeta _{-1}+\zeta _{0}^{2}=0$. The possible solutions are: 
\begin{equation}
\mathbf{C_{1}}:  \   \left(
\begin{array}{ccccc}
\sqrt{\frac{1+m}{3}}e^{i\theta _{2}}, & 0, & 0, & \sqrt{\frac{2-m}{3}}%
e^{i\theta _{-1}}, & 0%
\end{array}%
\right) ^{T},
\end{equation}%
with the energy $E_{C_1}/N=(2+m)q_0+\Tilde{c}_{0}/2+m^2\Tilde{c}_{1}/2$, and
\begin{equation}
\mathbf{C_{2}}: \   \left(
\begin{array}{ccccc}
0, & \sqrt{\frac{2+m}{3}}e^{i\theta _{1}}, & 0, & 0, & \sqrt{\frac{1-m}{3}}%
e^{i\theta _{-2}}%
\end{array}%
\right) ^{T},
\end{equation}%
with the energy $E_{C_2}/N=(2-m)q_0+\Tilde{c}_{0}/2+m^2\Tilde{c}_{1}/2$, and
\begin{equation}
\mathbf{C_{3}}:\  \left(
\begin{array}{ccccc}
\frac{m+2}{4}e^{i\chi}, & 0, & \frac{\sqrt4-m^2}{8},
& 0, & \frac{m-2}{4}e^{-i\chi}%
\end{array}%
\right) ^{T},
\end{equation}%
with the energy
\begin{align}  
\frac{E_{C_3}}{N}=&\frac{m^2+4}{2}q_0+\frac{\Tilde{c}_0}{2}+\frac{m^2\Tilde{c}_1}{2}+\frac{(m^2+16)\Tilde{c}_2}{128}   \notag\\
&-\frac{(m^2-4)\Tilde{c}_2}{64}J_0(8\mathcal{Q}). 
\end{align}  
Note that the spin structure of $\mathbf{C_{3}}$ is similar to that of polar phases $\mathbf{P_{3}}$, $\mathbf{P_{4}}$, and $\mathbf{P_{5}}$. However, for the cyclic state $\mathbf{C_{3}}$, the interaction coefficient $\Tilde{c}_2$ does not enter into the wavefunction but change the energy of the state.

\subsection{Broken-axisymmetry phase}

The characteristic of broken-axisymmetry phase is $\langle {F}_{+}\rangle \neq 0$ which may arise from the competition between the ferromagnetism $\Tilde{c}_1<0$ and the quadratic Zeeman energy $q_0$~\cite{Murata}.
With the constraints in Eq.~(\ref{constraints}), we find following possible broken-axisymmetry solutions.
\begin{equation}
\mathbf{BA_{1}}:\  \left(
\begin{array}{ccccc}
\sqrt{m-1}e^{i\theta_2}, & \sqrt{2-m}e^{i\theta_1}, & 0,
& 0, & 0%
\end{array}%
\right) ^{T},
\end{equation} 
with the energy
\begin{align}  
\frac{E_{BA_1}}{N}=(3m-2)q_0+\frac{\Tilde{c}_0}{2} +\frac{9m\Tilde{c}_1}{2}-4\Tilde{c}_1. 
\end{align}
\begin{equation}
\mathbf{BA_{2}}:\  \left(
\begin{array}{ccccc}
0, & \sqrt{m}e^{i\theta_1}, & \sqrt{1-m}e^{i\theta_0},
& 0, & 0%
\end{array}%
\right) ^{T},
\end{equation}
with the energy
\begin{align}  
\frac{E_{BA_2}}{N}=mq_0+\frac{\Tilde{c}_0}{2}+\frac{(m-1)^2}{2}\Tilde{c}_2+\left(3m-\frac{5m^2}{2}\right)\Tilde{c}_1  \notag 
\end{align} 
\begin{equation}
\mathbf{BA_{3}}:\  \left(
\begin{array}{ccccc}
0, & \frac{1+m}{2}e^{i\theta_1}, & \sqrt{\frac{1-m^2}{2}}e^{i\theta_0},
& \frac{1-m}{2}e^{i\theta_{-1}}, & 0%
\end{array}%
\right) ^{T},
\end{equation}
with the energy
\begin{align}  
\frac{E_{BA_3}}{N}&=\frac{1+m^2}{2}q_0+\frac{\Tilde{c}_0}{2} +\left[\frac{3(1-m^4)}{2}+m^2\right]\Tilde{c}_1  \notag \\ 
&+\frac{(1-m^2)^2}{4}\Tilde{c}_2+\frac{(3\Tilde{c}_1-\Tilde{c}_2 )(1-m^2)^2}{4}J_0(2\mathcal{Q}). \notag 
\end{align}
The above three states exist in a finite regime of $m$: $\mathbf{BA_{1}}$ in $1<m<2$, $\mathbf{BA_{2}}$ in $0<m<1$, and  $\mathbf{BA_{3}}$ in $-1<m<1$. Arbitrary phases appear in wavefunctions, leading to $\langle {F}_{+}\rangle $ depending on these phases, which becomes the signature of broken-axisymmetry states~\cite{Murata}. We also find another state that dose not depend on $m$.
\begin{equation}
\mathbf{BA_{4}}:\  \left(
\begin{array}{ccccc}
0, & \sqrt{\frac{1-\sigma^2}{2}}e^{i\theta_1}, & \sigma e^{i\theta_0},
& \sqrt{\frac{1-\sigma^2}{2}}e^{i\theta_{-1}}, & 0%
\end{array}%
\right) ^{T},
\end{equation}
with 
\begin{equation}
\sigma =\sqrt{ \frac{1}{2} - \frac{ q_0}{2(3\Tilde{c}_1-\Tilde{c}_2)[1+J_0(2\mathcal{Q})]}},
\label{sigma}
\end{equation}
and its energy is
\begin{align}
\frac{E_{BA_4}}{N} =& \frac{q_0}{2}+\frac{ \Tilde{c}_{0}}{2}+\frac{ \Tilde{c}_{2}}{2}-\frac{q
_0^2}{4(3\Tilde{c}_{1}-\Tilde{c}_{2})[1+J_0(2\mathcal{Q}) ]}  \notag \\
&+\frac{(3\Tilde{c}_{1}-\Tilde{c}_{2})[1+J_0{(2\mathcal{Q})}]}{4}.
\end{align}


\begin{figure}[t]
{\includegraphics[width = 0.3\textwidth]{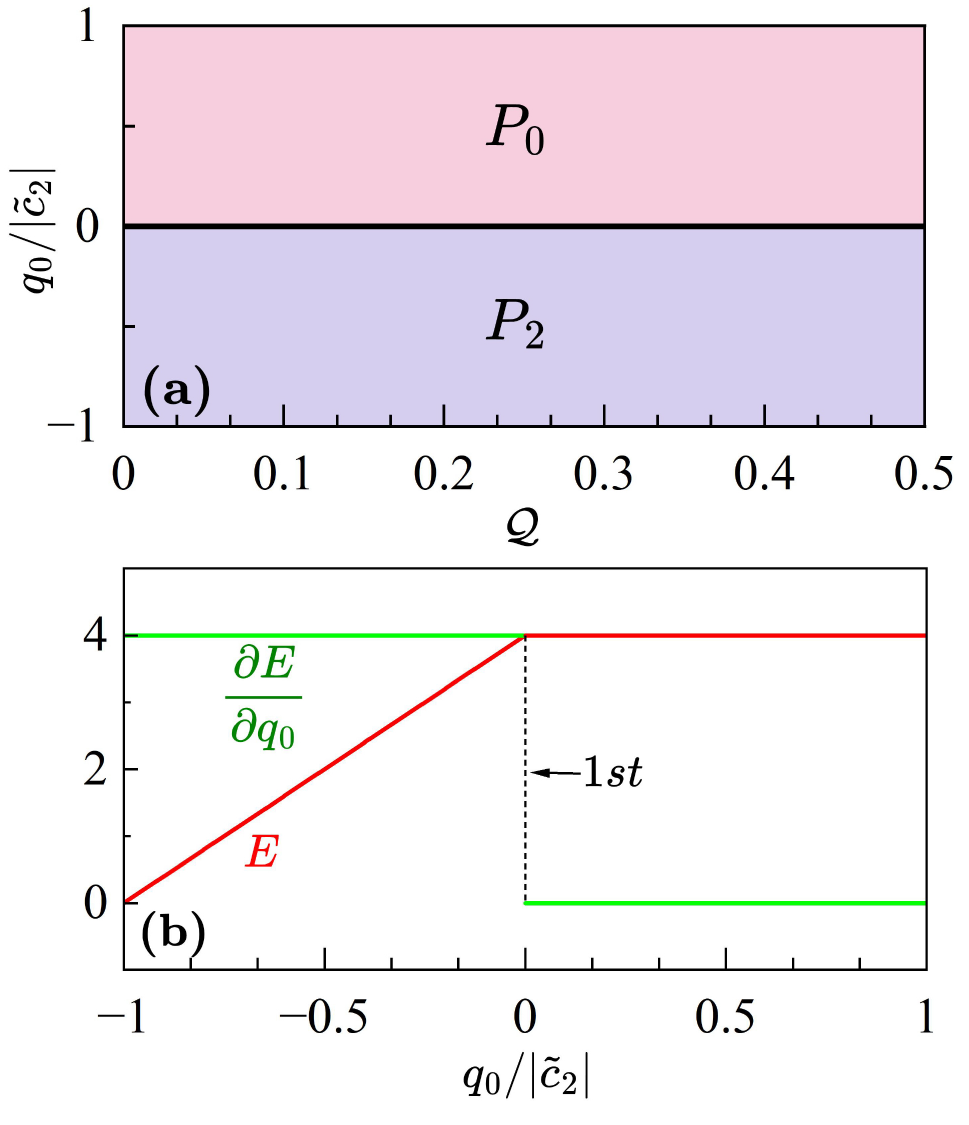}}
\caption{Ground-state phase diagram  for the anti-ferromagnetism $\Tilde{c}_1>0$ with $\Tilde{c}_2<0$ and $\langle F_{z}\rangle=m=0$. (a) Diagram in the parameter space of periodic driving $(q_0,\mathcal{Q}=Q/ (\hbar \omega))$. (b) Evolution of the energy and the derivative of energy as a function of $q_0$ at $\mathcal{Q}=0.2$. Other parameters are the same as  Fig.~\ref{fig1}.}     
\label{fig2}
\end{figure}

\section{ Ground-state phase diagrams}  
\label{diagram}

Once possible states are known, it would be interesting to show how they participate to ground-state phase diagram in the parameter space of periodic driving $(q_0,\mathcal{Q})$. For convenience, we scale all energy terms by $\left\vert \Tilde{c}_{2}\right\vert $ in the following study.
\subsection{Anti-ferromagnetism}

The spin-spin interaction $\Tilde{c}_{1}>0$ is the anti-ferromagnetism.  The complete minimization of ${\Tilde{c}_{1}}\langle \mathbf{F}\rangle ^{2}  $ in the energy functional would like to have $\langle {F}_{+}\rangle =0$ and $\langle F_{z}\rangle=0$, which motivates us to consider the constraint $m=0$.  With $m=0$, polar states and cyclic states may participate to phase diagram.  As the spin-singlet scattering $\Tilde{c}_2$ may play an important role in these possible states, we consider different ground-state phase diagrams by varying the sign of $\Tilde{c}_2$.

Fig.~\ref{fig1} demonstrates a ground-state phase diagram of the anti-ferromagnetism for a positive $\Tilde{c}_2$. Features of this phase diagram are addressed as follows.

(i) As shown in Fig.~\ref{fig1}(a), in the absence of the modulation, $\mathcal{Q}=0$, ground state is in the polar phase. When $|q_0|$ becomes large, the polar phase belongs to eigenstates of $F_z^2$. In order to minimize the energy of $q_0 \langle
F_{z}^{2}\rangle$, when $q_0>\Tilde{c}_2/2$ ground state is  $\mathbf{P_{0}}$ which is the eigenstate of $F_z^2$ with the eigenvalue $0$; when $q_0<-\Tilde{c}_2/2$ ground state is $\mathbf{P_{2}}$ which is the eigenstate of $F_z^2$ with the eigenvalue $4$. For the polar state $\mathbf{P_{4}}$, $\gamma=\sqrt{(2q_0+\Tilde{c}_2)/2\Tilde{c}_2} $ and $\sqrt{(1-\gamma^2)/2}=\sqrt{(\Tilde{c}_2-2q_0)/4\Tilde{c}_2}$, the existence of which requires $-\Tilde{c}_2/2 < q_0 < \Tilde{c}_2/2$.  Since the energy of $\mathbf{P_{4}}$ is always lower than $\mathbf{P_{0}}$ and $\mathbf{P_{2}}$, $\mathbf{P_{4}}$ becomes ground state in the region of $-\Tilde{c}_2/2 < q_0 < \Tilde{c}_2/2$. Fig.~\ref{fig1}(b) describes the evolution of the energy as a function of $q_0$. From the derivative curve (the green line), we identify the existence of the second-order phase transitions between $\mathbf{P_{2}}$ and $\mathbf{P_{4}}$ at $q=-\Tilde{c}_2/2$ and between $\mathbf{P_{4}}$ and $\mathbf{P_{0}}$ at $q=\Tilde{c}_2/2$.

(ii) In Fig.~\ref{fig1}(a), as increasing $\mathcal{Q}$ from 0, a new ground state of cyclic phase $\mathbf{C_{1}}$ appears locating around the center of $\mathbf{P_{4}}$ region and therefore splits the $\mathbf{P_{4}}$ region into two areas. The energy difference $E_{P_4}-E_{C_1} = \Tilde{c}_2(1-J_0(8\mathcal{Q})/4-4q_0^2/(\Tilde{c}_2(1+J_0(8\mathcal{Q}) ) $. In the region of $ - \Tilde{c}_2 \sqrt{ 1-J_0(8\mathcal{Q}) }/4 < q_0 <  \Tilde{c}_2 \sqrt{ 1-J_0(8\mathcal{Q}) }/4 $, $E_{P_4}-E_{C_1}>0$, indicating that $\mathbf{C_{1}}$ is the preferable ground state. The boundaries of $\mathbf{C_{1}}$ lie at  $ q_0 = \pm   \Tilde{c}_2 \sqrt{ 1-J_0(8\mathcal{Q}) }/4 $ and consequently extend with the crease of $\mathcal{Q}$. Meanwhile, the existence of $\mathbf{P_{4}}$ requires  $ - \Tilde{c}_2  (1+J_0(8\mathcal{Q}) )/4 < q_0 <  \Tilde{c}_2  (1+J_0(8\mathcal{Q}) )/4 $. Its boundaries $ q_0=\pm \Tilde{c}_2  (1+J_0(8\mathcal{Q}) )/4$ shrink with the increase of $\mathcal{Q}$. The extension and shrink of these two kinds of boundaries finally lead to the $\mathbf{P_{4}}$ regions shrink into two points at $\mathcal{Q}=0.3006$. The physical reason for the disappearance of the $\mathbf{P_{4}}$ phase is that at  $\mathcal{Q}=0.3006$, $J_0(8\mathcal{Q})=0$ leads to the boundaries of $\mathbf{C_{1}}$ and $\mathbf{P_{4}}$ phases becoming equal. 

The evolution of the energy as a function of $q_0$ at $\mathcal{Q}=0.15$ and $\mathcal{Q}=0.3006$ is demonstrated in Figs.~\ref{fig1}(c) and~\ref{fig1}(d) respectively. The calculation of the derivative of energy (green lines) indicates that both the transitions between $\mathbf{C_{1}}$ and $\mathbf{P_{4}}$ and the transitions between $\mathbf{C_{1}}$ and $\mathbf{P_{0}}$ are first order.

(iii) As further increasing $\mathcal{Q}$ from $\mathcal{Q}=0.3006$, the other new polar phase $\mathbf{P_{3}}$ appears at the edges of $\mathbf{C_{1}}$ region [see Fig.~\ref{fig1}(a)]. Physically, $\mathbf{P_{3}}$ state is closely similar to $\mathbf{P_{4}}$ but with the opposite sign in the front of $J_0(8\mathcal{Q})$. When $\mathcal{Q}>0.3006$, $J_0(8\mathcal{Q})<0$. It seems that $\mathbf{P_{3}}$ state takes over the region of $\mathbf{P_{4}}$ due to negative values of $J_0(8\mathcal{Q})$. The derivative of energy as a function of $q_0$ shown in Fig.~\ref{fig1}(e) illustrates that the transitions between  $\mathbf{C_{1}}$ and $\mathbf{P_{3}}$ are first order and the transitions between  $\mathbf{P_{3}}$ and $\mathbf{P_{0}}$ and between $\mathbf{P_{3}}$ and $\mathbf{P_{2}}$ are second order, providing a further information of the similarity of $\mathbf{P_{3}}$ and $\mathbf{P_{4}}$ phases.

Figure~\ref{fig2} presents a phase diagram for parameter identical to those in Fig.~\ref{fig1}, but with $\Tilde{c}_2 <0 $.  In contrast to the previous case, this diagram is remarkably simple, comprising only the $\mathbf{P_{0}}$ and $\mathbf{P_{2}}$ phases [see Fig.~\ref{fig2}(a)]. Furthermore, the driving has no effect on the phase boundaries, indicating that the Floquet and conventional spinor BECs are identical for this parameter set. The energy diagram in Fig.~\ref{fig2}(b) confirms that the transition between these phases is first order.


\begin{figure}[t]
{\includegraphics[width = 0.499\textwidth]{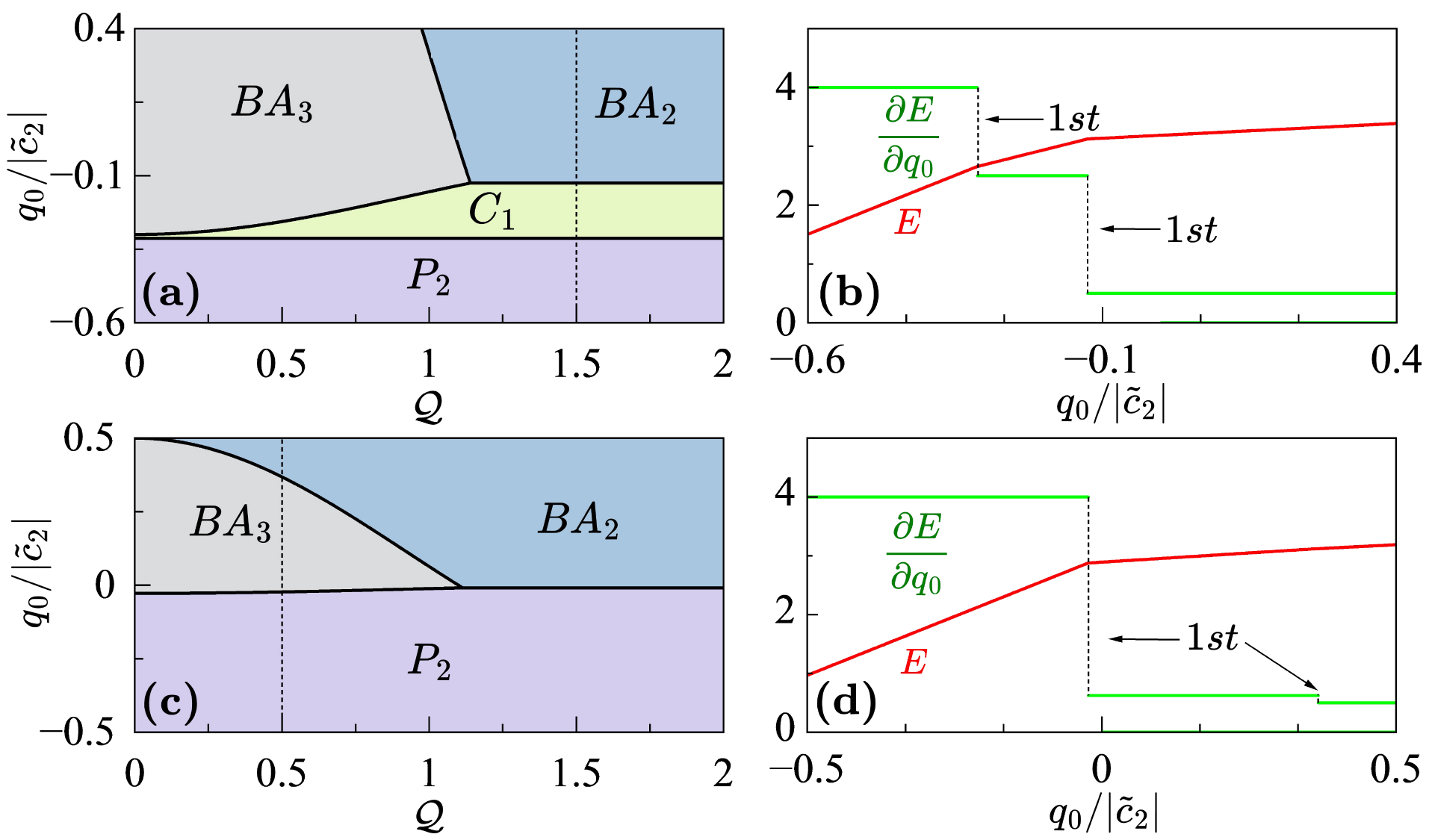}}
\caption{Ground-state phase diagram  for the ferromagnetism $\Tilde{c}_1<0$ with $\langle F_{z}\rangle=m=0.5$.
(a) Diagram in the parameter space of periodic driving $(q_0,\mathcal{Q}=Q/ (\hbar \omega))$ for $\Tilde{c}_2>0$.  The energy $E$ (red line) and derivative of energy $\partial E/\partial q_0$ (green line) along the vertical dashed line at $\mathcal{Q}=1.5$ are demonstrated in (b).  (c) Diagram for $\Tilde{c}_2<0$. The energy and derivative of energy  along the vertical dashed line at $\mathcal{Q}=0.5$ are shown in (d).  Other parameters are $ \Tilde{c}_{0}=7\left\vert
\Tilde{c}_{2}\right\vert $ and $\Tilde{c}_1=-0.5\left\vert
\Tilde{c}_{2}\right\vert $.}   
\label{fig3}
\end{figure}

\subsection{Ferromagnetism}

The spin-spin interaction $\Tilde{c}_{1}<0$ is the ferromagnetism.  The minimization of ${\Tilde{c}_{1}}\langle \mathbf{F}\rangle ^{2}  $ prefers to $\langle {F}_{+}\rangle \neq 0$ and $\langle F_{z}\rangle \neq 0$, which motivates us to study ground states with $m\neq 0$. For concrete, we consider an example $m=0.5$. In this case,  $\mathbf{BA_{2}}$, $\mathbf{BA_{3}}$ and $\mathbf{BA_{4}}$ can exist. 

 Fig.~\ref{fig3}(a) depicts a ground-state phase diagram of the ferromagnetism for a positive $\Tilde{c}_2$. (i) In the absence of modulation, $\mathcal{Q}=0$, the ground state resides in the  $\mathbf{P_{2}}$ phase for large negative values of $q_0 $. For large negative values of $q_0 $, the quadratic Zeeman energy $q_0 \langle
F_{z}^{2}\rangle$ completely dominates over other energy terms in the energy functional in Eq.~(\ref{Energyfunctional}) and it chooses the eigenstate of $F_z^2$ with the largest eigenvalue (which is $\mathbf{P_{2}}$ state) as ground state. Above the $\mathbf{P_{2}}$ phase, $\mathbf{BA_{3}}$ phase occupies a large area [see Fig.~\ref{fig3}(a)]. Between $\mathbf{P_{2}}$ and $\mathbf{BA_{3}}$ phases there is a very narrow occupation of the $\mathbf{C_{1}}$ state which is almost invisible in Fig.~\ref{fig3}(a). (ii) As increasing the modulation $\mathcal{Q}$ from 0, the $\mathbf{C_{1}}$ region expands in the way that the upper boundary between $\mathbf{C_{1}}$ and $\mathbf{BA_{3}}$ phases extends into $\mathbf{BA_{3}}$ phase but the lower boundary between the  $\mathbf{C_{1}}$ and $\mathbf{P_{2}}$ phases keeps as a constant. Since energies of  $\mathbf{C_{1}}$ and $\mathbf{P_{2}}$ do not depend on the modulation,  a finite $\mathcal{Q}$ can not modify their phase boundary. However, the energy of  $\mathbf{BA_{3}}$ state explicitly depends on the modulation via Bessel function $ E_{BA_3 } =  5q_0/8 +45\Tilde{c}_{2}/16-45\Tilde{c}_{2}J_0(2\mathcal{Q}) /128 $. According to $ E_{C_1 }-E_{BA_3 } =  15q_0/8 +27\Tilde{c}_{2}/128+45\Tilde{c}_{2}J_0(2\mathcal{Q}) /128 $,  the boundary between $\mathbf{C_{1}}$ and $\mathbf{BA_{3}}$ phases lies at $q_0=-9\Tilde{c}_{2}/80-3\Tilde{c}_{2} J_0(2\mathcal{Q}) /16$. Therefore, the boundary increases as the increase of $\mathcal{Q}$. (iii) On the other hand, since the increase of $\mathcal{Q}$ can enhance the energy of  $\mathbf{BA_{3}}$ phase, after critical $\mathcal{Q}$, the $\mathbf{BA_{3}}$ phase is not preferable as ground state and is taken over by the $\mathbf{BA_{2}}$ phase [see Fig.~\ref{fig3}(a)]. Moreover, the energy of  $\mathbf{BA_{2}}$ phase does not depend on the modulation. Therefore, the boundary between $\mathbf{BA_{2}}$ and $\mathbf{C_{1}}$ phases does not change as increasing $\mathcal{Q}$. Finally, phase transitions between $\mathbf{P_{2}}$ and $\mathbf{C_{1}}$ phases and between $\mathbf{BA_{2}}$ and $\mathbf{C_{1}}$ phases are first order as indicated from the evolution of energies shown in Fig.~\ref{fig3}(b). 

Fig.~\ref{fig3}(c) demonstrates a ground-state phase diagram of the ferromagnetism for a negative $\Tilde{c}_2$. The phase-diagram structure is similar to that in Fig.~\ref{fig3}(a), but  the $\mathbf{C_{1}}$ phase is no longer present. The minimization of $\Tilde{c}_{2} \left\vert \Theta \right\vert ^{2} $ with  $\Tilde{c}_2<0$ prefers to $\Theta \neq 0 $. Therefore, it is expected that cyclic phases can not distribute in phase diagram in Fig.~\ref{fig3}(c). It is interesting to note that phase transitions between $\mathbf{P_{2}}$ and $\mathbf{BA_{3}}$ phases and between $\mathbf{BA_{3}}$ and $\mathbf{BA_{2}}$ phases are still keeping as first order [see Fig.~\ref{fig3}(d)].

We want to emphasis that the value of $m$ has a important effect on ground-state phase diagrams. For a larger value, for example, $m=1.5$, only $\mathbf{BA_{1}}$ state can exist. For $\Tilde{c}_2>0$, according to the structure in Fig.~\ref{fig3}(a), ground states include $\mathbf{BA_{1}}$, $\mathbf{C_{1}}$ and $\mathbf{P_{2}}$ phases. Since energies of these phases do not depend on the modulation, Floquet-engineered interactions have no effect to induce phase re-distributions. For $\Tilde{c}_2<0$, only $\mathbf{BA_{1}}$ and $\mathbf{P_{2}}$ are left as ground state. Similarly, adjusting $\mathcal{Q}$ can not change phase distribution of  $\mathbf{BA_{1}}$ and $\mathbf{P_{2}}$.

Finally, we discuss the possible experimental observation of the cyclic phase, which distinguishes spin-2 spinor BECs from their spin-1 counterparts. However, as shown in Fig.~\ref{fig1}(a) and Fig.~\ref{fig3}(a), the cyclic phase either cannot become the ground state or exists only in an extremely narrow region of $q_0$ in the absence of driving ($\mathcal{Q}=0$). To date, the cyclic phase remains challenging to observe experimentally \cite{yuki12,Schmaljohann2004,ho00}. Figures \ref{fig1}(a) and \ref{fig3}(a) demonstrate that Floquet engineering can substantially expand its region of existence as the ground state. Therefore, spin-2 Floquet spinor BECs provide a promising and experimentally accessible platform for observing the cyclic phase.

\section{Conclusions}
\label{conclusion}

We propose to implement a spin-2 Floquet spinor BEC by Floquet engineering of the quadratic Zeeman energy.   Coupling strengths of all spin-flip interacting processes that conserve angular momentum in spin-2 Floquet spinor BECs are effectively modulated by Bessel functions which depend on driving parameters.  By changing driving parameters, interaction coefficients of all these processes can be relatively tuned. Considering experimental accessibility to adjust the quadratic Zeeman energy, the proposed Floquet system provides a significant platform with tunable spin-dependent interactions. As shown in Ref.~\cite{Guan2025} in a spin-1 spinor BEC, the spin-flip interacting coupling has an important effect on the spin-mixing dynamics, spin-2 Floquet spinor BECs are another ideal systems to explore spin dynamics. 

We find that the Floquet-engineered interactions can significantly influence possible  states in a homogeneous system. In particular, they support new states that are absent in conventional spin-2 spinor Bose-Einstein condensates. Ground-state phase diagrams are analyzed for fixed interaction coefficients, illustrating the distribution of possible ground states in the space of driving parameters.  A key feature of the ground-state phase diagrams is that Floquet spinor systems can facilitate the experimental observation of the cyclic phase.

\section{Acknowledgments}

This work is supported by the National Natural Science
Foundation of China (NSFC) under Grants No. 12374247 and
No. 11974235, as well as by the Shanghai Municipal Science and Technology Major Project (Grant No. 2019SHZDZX01-ZX04).

\section{Appendix}
\label{appendix}

We derive the similar Floquet engineered Hamiltonian $\hat{H}_\text{eff}$ in Eq.~(\ref{effectiveH}) using the high-frequency expansion~\cite{gold14}. This approach emphasizes the importance of non-communication between the quadratic Zeeman energy and the interactions.  

To proceed, we first rewrite the Hamiltonian in Eq.~(\ref{Orignal}) as~\cite{gold14}
\begin{equation}
    \hat{H}= \hat{H}_0 + \hat{V}\cos(\omega t),
\end{equation}
with
\begin{align}
\hat{H}_0= & \int d\mathbf{r} \hat{\psi}^{\dagger }\left( -\frac{\hbar 
^{2}\triangledown ^{2}}{2M}\right) \hat{\psi} + \int d\mathbf{r} \hat{\psi}^{\dagger }\left( q_0F_{z}^{2}\right) \hat{\psi} \notag \\
&+  \frac{c_{0}}{2}\int d\mathbf{r}\left( \hat{\psi}^{\dagger }%
\hat{\psi}\right) ^{2}  
+\frac{c_{1}}{2}\int d\mathbf{r}\left( \hat{\psi}^{\dagger }%
\mathbf{{F}}\hat{\psi}\right) \cdot \left( \hat{\psi}^{\dagger }\mathbf{%
{F}}\hat{\psi}\right)  \notag \\
&+
\frac{c_{2}}{2}\int d\mathbf{r}\left( \hat{\psi}^{T}{A%
}\hat{\psi}\right) ^{\dagger }\left( \hat{\psi}^{T}{A}\hat{\psi}%
\right), \notag
\end{align}
and
\begin{equation}
    \hat{V}= \int d\mathbf{r} \hat{\psi}^{\dagger }\left( QF_{z}^{2}\right) \hat{\psi}. \notag
\end{equation}
The evolution operator is $U(T,0)=\mathcal{T}\exp(-i/\hbar \int_0^T dt\hat{H} )$ with $T=2\pi/ \omega$ and $\mathcal{T}$ denoting the time-ordering. The Floquet Hamiltonian $\hat{H}_\text{eff}$ is defined as~\cite{eck15}
\begin{equation}
   \mathcal{T}e^{-i/\hbar \int_0^T dt\hat{H} } = e^{-i/\hbar \hat{H}_\text{eff} T}.
\end{equation}
Assume $\omega $ is high, and $Q/\omega$ shall be a small quantity. Using high-frequency expansion, we get~\cite{gold14},
\begin{equation}
\hat{H}_\text{eff}=\hat{H}_{0}+\frac{ 1}{4 (\hbar \omega) ^{2}}\left[\left[\hat{V},\hat{H}
_{0}\right],\hat{V}\right]+\mathcal{O}(Q^3/  \omega ^{3}).  \label{perturabation}
\end{equation}%
The commutator is calculated
\begin{align}
& \left[\left[\hat{V},\hat{H}_{0}\right],\hat{V}\right]  \notag \\
&=-4Q^2c_{1}  \int d\mathbf{r} \left( 3\hat{\psi}%
_{0}^{\dagger }\hat{\psi}_{0}^{\dagger }\hat{\psi}_{1}\hat{\psi}_{-1}+\sqrt{6%
}\hat{\psi}_{1}^{\dagger }\hat{\psi}_{1}^{\dagger }\hat{\psi}_{2}\hat{\psi}%
_{0}\right.  \notag \\
&  +\sqrt{6}\hat{\psi}_{-1}^{\dagger }\hat{\psi}%
_{-1}^{\dagger }\hat{\psi}_{0}\hat{\psi}_{-2} +18\hat{\psi}_{-1}^{\dagger }\hat{\psi}_{1}^{\dagger }\hat{\psi}_{2}\hat{%
\psi}_{-2} \notag \\
&\left. +4\sqrt{6}\hat{\psi}_{-1}^{\dagger }\hat{\psi}_{0}^{\dagger }\hat{%
\psi}_{1}\hat{\psi}_{-2} +4\sqrt{6}\hat{\psi}_{0}^{\dagger }\hat{\psi}_{1}^{\dagger }\hat{%
\psi}_{2}\hat{\psi}_{-1}+\text{H.c.}\right)  \notag   \\
&-4Q^2c_{2} \int d\mathbf{r}\left( - \hat{\psi}_{0}^{\dagger }\hat{\psi}_{0}^{\dagger }\hat{\psi}_{1}%
\hat{\psi}_{-1}+16\hat{\psi}_{0}^{\dagger }\hat{%
\psi}_{0}^{\dagger }\hat{\psi}_{2}\hat{\psi}_{-2} \right.  \notag \\
&\left. -18\hat{\psi}_{-1}^{\dagger }\hat{\psi}%
_{1}^{\dagger }\hat{\psi}_{2}\hat{\psi}_{-2}+\text{H.c.}\right),
\end{align}
which is a signature of the non-communication between the quadratic Zeeman energy and the interactions. By substituting the above result into Eq.~\eqref{perturabation}, we obtain the Floquet Hamiltonian,
\begin{widetext}
\begin{align}
\hat{H}_\text{eff} =&\hat{H}_{0}
-\frac{c_1}{2}  \left( \frac{Q}{\hbar \omega}\right)^2 \int d\mathbf{r}  \left( 6\hat{\psi}%
_{0}^{\dagger }\hat{\psi}_{0}^{\dagger }\hat{\psi}_{1}\hat{\psi}_{-1}+2\sqrt{6%
}\hat{\psi}_{1}^{\dagger }\hat{\psi}_{1}^{\dagger }\hat{\psi}_{2}\hat{\psi}%
_{0}  +2\sqrt{6}\hat{\psi}_{-1}^{\dagger }\hat{\psi}%
_{-1}^{\dagger }\hat{\psi}_{0}\hat{\psi}_{-2}   +36\hat{\psi}_{-1}^{\dagger }\hat{\psi}_{1}^{\dagger }\hat{\psi}_{2}\hat{%
\psi}_{-2}  \right. \notag \\
&\left. +8\sqrt{6}\hat{\psi}_{-1}^{\dagger }\hat{\psi}_{0}^{\dagger }\hat{%
\psi}_{1}\hat{\psi}_{-2} +8\sqrt{6}\hat{\psi}_{0}^{\dagger }\hat{\psi}_{1}^{\dagger }\hat{%
\psi}_{2}\hat{\psi}_{-1}+\text{H.c.} \right)  - \frac{c_2}{2}  \left( \frac{Q}{\hbar \omega}\right)^2\int d\mathbf{r} \left(  
32\hat{\psi}_{0}^{\dagger }\hat{%
\psi}_{0}^{\dagger }\hat{\psi}_{2}\hat{\psi}_{-2}  -36\hat{\psi}_{-1}^{\dagger }\hat{\psi}%
_{1}^{\dagger }\hat{\psi}_{2}\hat{\psi}_{-2}
\right. \notag   \\
& \left.  -2\hat{\psi}_{0}^{\dagger }\hat{\psi}_{0}^{\dagger }\hat{\psi}_{1}%
\hat{\psi}_{-1}  +   \text{H.c.}\right),
\label{other}
\end{align}%
\end{widetext}
Note that there is a difference between the above Hamiltonian from the high-frequency expansion and the engineered one from unitary transformation and time average in  Eq.~(\ref{effectiveH}), as we only keep the second order of the perturbation of $Q/(\hbar \omega)$ in Eq.~(\ref{perturabation}). By expanding the Bessel function $J_0( jQ/(\hbar\omega) ) \approx 1-j^2Q^2/(4\hbar^2\omega^2)$ with the assumption $jQ/(\hbar\omega)$ being a small quantity, the effective Hamiltonian in Eq.~(\ref{effectiveH}) automatically turns to be the perturbed one in Eq.~(\ref{other}).

\end{document}